\documentclass[11pt,a4paper]{article}
\usepackage{jcappub}

\usepackage{bm}
\usepackage{amsfonts}
\usepackage{subfigure}

\newcommand{\be}{\begin{equation}}
\newcommand{\ee}{\end{equation}}
\newcommand{\bea}{\begin{eqnarray}}
\newcommand{\eea}{\end{eqnarray}}

\newcommand{\di}{\text{d}}

\begin{document}



\title{Getting leverage on inflation with a large photometric redshift survey}

\author[a]{Tobias Basse}
\author[b]{Jan Hamann}
\author[a,c]{Steen Hannestad}
\author[d]{Yvonne~Y.~Y.~Wong}

\affiliation[a]{Department of Physics and Astronomy,
 University of Aarhus, DK-8000 Aarhus C, Denmark}
\affiliation[b]{Theory Division, Physics Department,
CERN, CH-1211 Geneva 23, Switzerland} 
\affiliation[c]{Aarhus Institute of Advanced Studies,
 Aarhus University, DK-8000 Aarhus C, Denmark}
\affiliation[d]{School of Physics, 
The University of New South Wales, Sydney NSW 2052, Australia}

\emailAdd{basse@phys.au.dk, jan.hamann@cern.ch, sth@phys.au.dk, yvonne.y.wong@unsw.edu.au}

\abstract{
We assess the potential of a future large-volume photometric redshift survey to constrain observational inflationary parameters using three large-scale structure observables: 
the angular shear and galaxy power spectra, and the cluster mass function measured through weak lensing. When used in combination with Planck-like CMB measurements, 
we find that the spectral index~$n_{\rm s}$ can be constrained to a $1 \sigma$ precision of up to $0.0025$.
The sensitivity to the running of the spectral index can potentially improve to $0.0017$, roughly a factor of five better than the present $1\sigma$~constraint from Planck and auxiliary CMB data, allowing us to test the assumptions of the slow-roll scenario with unprecedented accuracy.
Interestingly, neither CMB+shear nor CMB+galaxy nor CMB+clusters {\it alone} can achieve this level of sensitivity; it is the combined power of all three probes that conspires to break the different parameter degeneracies inherent in each type of observations.  We make our forecast software publicly available via download or upon request from the authors.
}

\maketitle

\section{Introduction}
Cosmic inflation is currently the most compelling mechanism to explain the generation of the primordial fluctuations.  Precision measurements of anisotropies in the cosmic microwave background (CMB) by the Planck~\cite{Ade:2013ktc} and WMAP~\cite{Bennett:2012zja} missions are consistent with adiabatic, Gaussian primordial scalar perturbations that are nearly scale-invariant~\cite{Ade:2013uln}, as predicted by the simplest slow-roll models of inflation. This picture is likewise compatible with the claimed detection of primordial gravitational waves via the $B$-mode signal by the BICEP-2 experiment~\cite{Ade:2014xna}.
  Interestingly, the data do show a weak preference for a negative running of the scalar spectral index, $\mathrm{d} n_\mathrm{s}/\mathrm{d}\ln k = -0.013 \pm 0.009$ at~$1\sigma$~\cite{Ade:2013zuv}, which if true would imply the presence of non-negligible third- and higher-order derivatives of the inflaton potential~\cite{Easther:2006tv}.
These are not a generic inflationary prediction, and, if ever established at high statistical significance, would rule out a large number of simple candidate models. 

To further the constraints on and/or the detection of~$\mathrm{d} n_\mathrm{s}/\mathrm{d} \ln k$, it would be extremely helpful to increase the lever arm in wavenumbers $k$ by measuring cosmic perturbations on even smaller length scales.  Unfortunately, the CMB's sensitivity here is limited by a combination of the primary signal's suppression due to Silk damping, and the increasing importance of foreground anisotropies.
Density perturbations of the matter field at later times are however not subject to these limitations, and with several  large-volume photometric redshift surveys such as the Large Synoptic Survey Telescope (LSST)~\cite{Abell:2009aa} and the Euclid mission~\cite{Laureijs:2011mu} mapping out a significant fraction of the local universe in the coming decade, it is timely to address the question of to what extent the information about smaller scale perturbations gained by these surveys will further our understanding of the physics governing the inflationary era.  

To this end, we consider the three most important cosmological observables of a future large-volume photometric redshift surveys---the angular shear power spectrum, the angular galaxy power spectrum, and the cluster mass function---and their sensitivity to inflationary parameters.
We generate mock data for the said cosmological observables given the survey specifications, and then analyse the mock data using Markov Chain Monte Carlo (MCMC) methods.  Our main analysis tool is based on the publicly available \texttt{CosmoMC}~\cite{Lewis:2002ah} code, 
with modifications first described in~\cite{Hamann:2012fe} and later extended in~\cite{Basse:2013zua}. 
Besides reporting our forecast of the expected sensitivity of future photometric redshift surveys to inflationary parameters, 
we would also like to take this opportunity to publicly release the code for general use.\footnote{A copy of the forecast software based on the July 2014 version of \texttt{CosmoMC} can be downloaded from \texttt{http://jhamann.web.cern.ch/jhamann/simdata/simdata.tar.gz}.}

The paper is organised as follows. In section~\ref{sec:pps} we briefly describe the parameterisation of primordial fluctuations used in the analysis.  
Section~\ref{sec:data} reviews the cosmological observables and recapitulates the technical details of the mock data generation from~\cite{Hamann:2012fe,Basse:2013zua}, while
section~\ref{sec:forecast} describes the forecast model parameter space and the likelihood functions.  Our results are presented in section~\ref{sec:results}, followed by  more in-depth discussions pertaining to inflationary physics in section~\ref{sec:sloro}.  Section~\ref{sec:conc} contains our conclusions.


\section{Parametrisation of the primordial power spectra \label{sec:pps}}

We parameterise the primordial scalar and tensor power spectra in the usual way. Starting with
\begin{equation}
\begin{aligned}
{\cal P}_{\cal R} & =  A_\mathrm{s} \left(\frac{k}{k_\mathrm{p}}\right)^{n_\mathrm{s}(k)-1}, \\
{\cal P}_\mathrm{t} & =  A_\mathrm{t} \left(\frac{k}{k_\mathrm{p}}\right)^{n_\mathrm{t}(k)}, \\
\end{aligned}
\end{equation}
we Taylor-expand $n_{\mathrm{s},\mathrm{t}}(k)$  to second order around the pivot scale~$k_\mathrm{p}$, so that
\begin{equation}
\begin{aligned}
{\cal P}_{\cal R} & =  A_\mathrm{s} \left(\frac{k}{k_\mathrm{p}}\right)^{n_\mathrm{s}(k_{\rm p})-1+\frac{1}{2} \mathrm{d}n_\mathrm{s}/\mathrm{d}\ln k \; \ln(k/k_\mathrm{p}) + ...}, \\
{\cal P}_\mathrm{t} & =  A_\mathrm{t} \left(\frac{k}{k_\mathrm{p}}\right)^{n_\mathrm{t}(k_{\rm p})+\frac{1}{2} \mathrm{d}n_\mathrm{t}/\mathrm{d}\ln k \; \ln(k/k_\mathrm{p}) + ...}.
\end{aligned}
\end{equation}
The choice of pivot scale is arbitrary.  In this work, we use $k_\mathrm{p} = 0.05$~Mpc$^{-1}$ following common convention.

It is also common to express the tensor fluctuation amplitude~$A_{\rm t}$ in terms of the tensor-to-scalar ratio~$r$,  defined as the ratio of tensor to scalar power at the pivot scale,
\begin{equation}
r \equiv \left. \frac{{\cal P}_{\cal R}}{{\cal P}_{\rm t}}  \right|_{k=k_{\rm p}} = \frac{A_\mathrm{t}}{A_\mathrm{s}}.
\end{equation}
Then, using the inflationary slow-roll consistency relation $r = -8 n_\mathrm{t}$, and neglecting any running that might be present in $n_{\rm t}(k)$, 
we are left with four parameters 
\begin{equation}
\Theta^{(\rm infl)} \equiv \left(A_\mathrm{s}, n_\mathrm{s}(k_\mathrm{p}), \mathrm{d} n_\mathrm{s}/\mathrm{d}\ln k(k_\mathrm{p}), r \right) 
\end{equation}
to describe the primordial scalar and tensor fluctuations.


\section{Cosmological observables and mock data generation}
\label{sec:data}

We consider four cosmological observables, CMB (temperature and polarisation), the angular cosmic shear power spectra, the angular galaxy power spectra, and 
the cluster mass function.  We briefly review below these observables and the generation of synthetic data thereof, 
but refer the reader to more detailed descriptions in~\cite{Hamann:2012fe} (CMB, shear, and galaxy) and~\cite{Basse:2013zua} (clusters).


\subsection{Shear and galaxy power spectra}

Surveys such as the LSST and Euclid are dedicated cosmic shear surveys that will measure the shapes and the positions of millions of galaxies.
The cosmic shear power spectrum will be a prime product of these measurements, but the galaxy power spectrum and its cross-correlation with cosmic shear will also come for free.  
Because photometric surveys are relatively poor at determining the radial positions of galaxies, it is more advantageous to work with angular power spectra~$C_\ell$, instead  of the full three-dimensional power spectra~$P(k)$.  (See, e.g.,~\cite{Audren:2012vy} for a parameter sensitivity forecast using the three-dimensional matter power spectrum.)
However, even in this case it is possible to group the observed galaxies into broad redshift bins.  Thus, the final observables are $C^{XY}_{\ell,ij}$, where $X,Y$ refer to either shear (s) or galaxies~(g), and $i,j$ to the redshift bin number of $X$ and $Y$ respectively.

Once the cosmological model has been specified, the angular power spectra~$C^{XY}_{\ell, ij}$ can be computed from theory using
\begin{equation}
\label{eq:cl}
  \mathcal{C}_{\ell,ij}^{XY} = 4 \pi \int \di \ln k \;  \mathcal{S}_{\ell,i}^X(k) \, \mathcal{S}_{\ell,j}^Y(k) \, {\mathcal P}_{\mathcal R}(k).
\end{equation}
Here,  $ {\mathcal P}_{\mathcal R}(k)$ is the dimensionless power spectrum of the primordial curvature perturbations~${\mathcal R}(k)$, and
the source functions for shear and galaxies read, respectively, 
\begin{equation}
\begin{aligned}
\label{eq:source1}
  \mathcal{S}_{\ell,i}^{\mathrm s} & =  -2 \ \sqrt{\frac{\ell (\ell^2 - 1) (\ell+2)}{4}} \int \di \chi \; j_\ell (k \chi) \, \mathcal{W}_i^{\rm s}(\chi)  \, T_\Psi(k,\eta_0 - \chi), \\
  \mathcal{S}_{\ell,i}^{\mathrm g} & =  \int \di \chi \; j_\ell (k \chi) \, \mathcal{W}_i^{\rm g} (\chi) \, T_\delta(k,\eta_0 - \chi) ,
\end{aligned}
\end{equation}
 where the integration variable $\chi$ is the comoving distance, $\eta_0$ is the conformal time today, and $j_\ell$ are the spherical Bessel functions. 

 The source functions incorporate different  transfer functions~$\mathcal{T}_{\Psi, \delta}$ of the metric perturbations $\Psi$ and  
  the matter density fluctuations~$\delta$, respectively~\cite{Hamann:2012fe}.
The window  functions $\mathcal{W}^X_i$ likewise differ, and are given by
\begin{equation}
\begin{aligned}
  \mathcal{W}_i^{\mathrm g}  ( \chi)& = \int_\chi^\infty \di \chi' \; \frac{\chi-\chi'}{\chi'\chi} \hat{n}_i(\chi'), \\
  \mathcal{W}_i^{\mathrm g}(\chi) & = b(k,\chi)  \hat{n}_i(\chi),
\end{aligned}
\end{equation}
with
\begin{equation}
\label{eq:galredshift}
  \hat{n}_i(\chi) = H(z) \hat{n}_i(z) = H(z) \frac{\di n/\di z(z)}{n_i} ,
  \end{equation}
where $\di n(z)/\di z$ is a survey-specific source galaxy distribution taken here to be of the form
\begin{equation}
{\rm d}n(z)/{\rm d}z \propto z^2 \exp(-(z/z_0)^\beta),
\label{eq:galdist}
\end{equation}  
and normalised to the surface density of source galaxies in bin $i$ via
\begin{equation}
\label{eq:ni}
 n_i \equiv \int_{\Delta z_i} \di z' \; \di n(z')/\di z'.
 \end{equation}
Distribution~(\ref{eq:galdist}) is a universal form for magnitude limited surveys~\cite{Kaiser:1996tp,Koo:1996ix,Hu:1999ek}. Here we use $\beta = 1$ and $z_0 = 0.3$,
so as to reproduce  Euclid's projected median survey redshift of $\sim 0.8$.

Finally, the galaxy window function $\mathcal{W}_i^{\mathrm g}$ also depends on the
galaxy bias $b(k,\chi)$ relating the galaxy number density fluctuations to the underlying matter density perturbations.


\subsubsection{Measurement errors\label{sec:errors}}

Even if observations were perfect, the construction of the galaxy and shear power spectra still suffer from a sample (cosmic) variance of $\Delta C_\ell/C_\ell = \sqrt{2/(2 \ell +1)}$. This irreducible uncertainty will be accounted for at the level of the likelihood function (section~\ref{sec:likelihood}).
Measurement-related uncertainties, however, must be factored into the mock data generation.
We discuss these briefly below.
 
 \paragraph{Shot noise\label{sec:shotnoise}}

The source density of galaxies determines the smallest angular separation (and thus the highest $\ell$) on the sky at which $C^{XY}_{\ell,ij}$ can be reliably measured. This finite surface density appears as a shot-noise term directly in the power spectrum
\begin{equation}
\label{eq:shotnoise}
\delta C_{{\rm noise},ij}^{XY} = \delta_{ij} \delta_{XY} (\Xi^{X}_{i})^2 n_i^{-2},
\end{equation}
which should be added to the primary power spectrum to obtain t the total observed power.  
Here,
$\Xi^{\rm g}_i=1$ for the galaxy power spectrum measurement,  $\Xi^{\rm s}_i = \langle \gamma^2 \rangle^{1/2}$ corresponds to the root-mean-square ellipticity of the source galaxies ($\langle \gamma^2 \rangle ^{1/2} \sim 0.35$ for Euclid~\cite{Basse:2013zua}), and
the Kronecker deltas $\delta_{ij}$ and $\delta_{XY}$ signify that the noise term contributes only to the auto-spectra.
The source galaxy surface densities $n_i$ in the redshift bin $i$ can be computed using equation~(\ref{eq:ni}).

\paragraph{Photometric redshift uncertainties}

We model the photometric redshift  uncertainty as a simple Gaussian error of standard deviation $\sigma(z) = 0.03 (1+z)$~\cite{Laureijs:2011mu} and with no bias.  
(We note that the Dark Energy Survey has already achieved~$\sigma(z) \sim 0.08$ reliably and with very little bias~\cite{Sanchez:2014rga}.)
This error can be folded into the window functions $\mathcal{W}^X_i$ as described in reference~\cite{Ma:2005rc}. 
However, we stress that, using no more than about ten redshift bins, the bin widths are typically much larger than $\sigma(z)$; increasing the number of redshift bins to beyond ten (and consequently decreasing the bin widths) also does not substantially improve the parameter sensitivies~\cite{Hamann:2012fe}.
Therefore, even if finally $\sigma(z)$ could only be controlled to $0.08$, it would have little impact on our forecasted parameter sensitivities.

In principle, redshift space distortions should also be taken into account.  However, because these are of order $10^3~{\rm km} \ {\rm s}^{-1}$, corresponding to $\delta z \sim 0.003$, they will be subdominant to the effect of the photometric redshift error at all but the smallest redshifts.  We therefore ignore them in our analysis.
Likewise we ignore possible errors in the shear measurements arising from intrinsic alignment and the shape measurement of source galaxies.


\subsubsection{Effective $k$-range of the shear and galaxy data}\label{sec:effectivek}

As we see from equation~(\ref{eq:cl}), the angular power spectrum $C_{\ell, ij}^{XY}$ is formed by integrating over $\ln k$ two source functions~$S_{\ell, i}^X(k)$ and the primordial curvature perturbation power spectrum~${\cal P}_{\cal R}(k)$.  
The role of the source functions, therefore, is to pick out those values of $k$ that are relevant for the multipole~$\ell$ in the redshift bins concerned.

\begin{figure}[t]
\center
\includegraphics[height=.66\textwidth,angle=0]{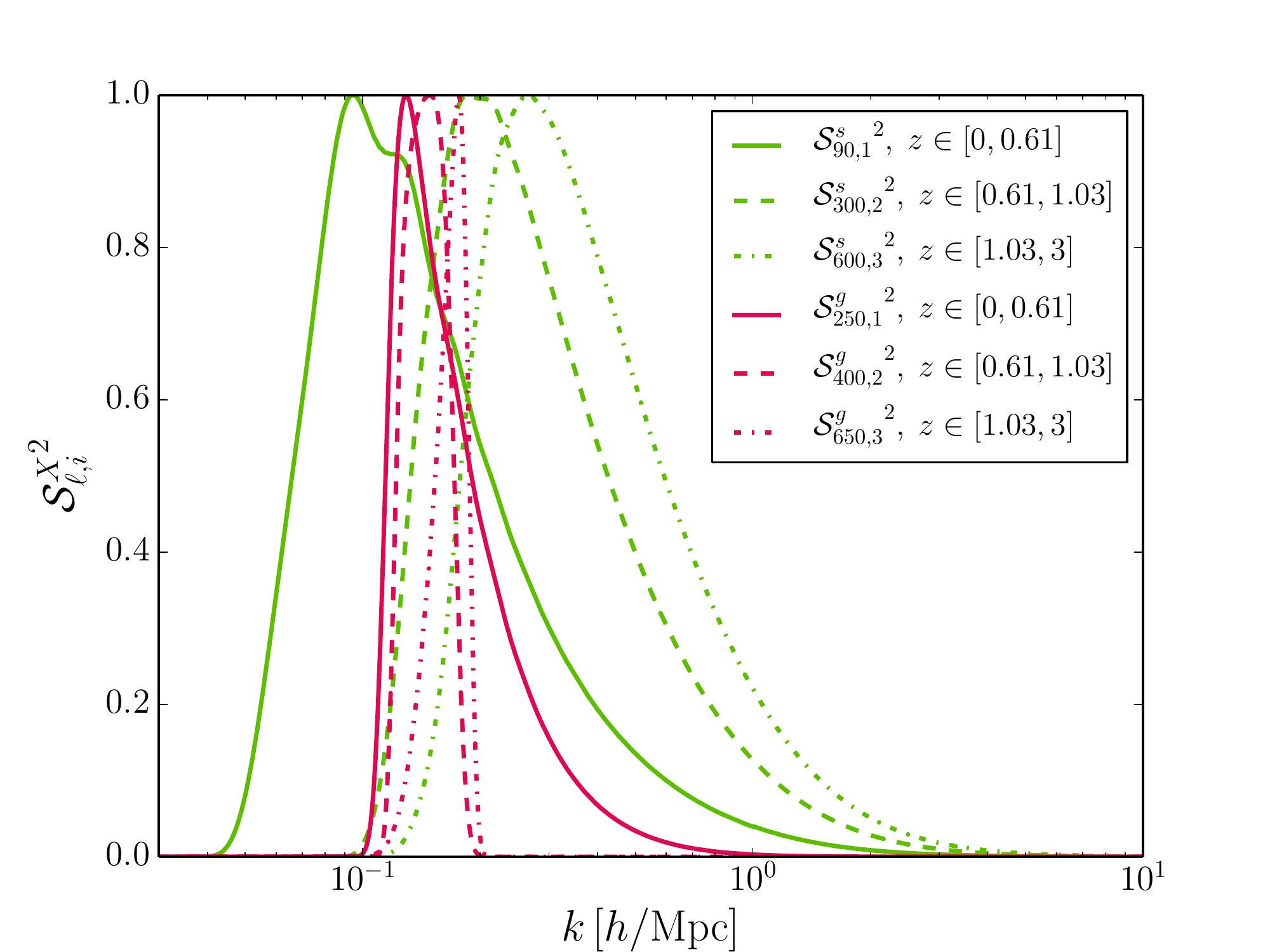}
\caption{The square of the shear and galaxy source functions, ${S^{\rm s}_{\ell,i}}^2(k)$ and ${S^{\rm g}_{\ell,i}}^2(k)$, in three redshift bins, as functions of $k$.  The $\ell$ value in each case has been chosen to be the highest accessible to that observable at the redshift concerned.
For shear this is the multipole at which the shot noise overtakes the signal, while for galaxy clustering this is determined by our non-linearity cut-off.\label{fig:source}}
\end{figure}

Figure~\ref{fig:source} shows the source functions for shear and galaxy clustering at the highest accessible multipole~$\ell$ in three redshift bins. Clearly, in every redshift bin, the shear source function consistently picks out higher $k$~values than the galaxy source function, even though the highest accessible~$\ell$ in the latter observable generally exceeds the former case.  This is because shear is a cumulative line-of-sight effect; while the galaxy source function selects  predominantly $k \sim \ell/\chi(z_i)$, where $z_i$ is, say, the mean redshift of the $i$-bin, for the same $\ell$ the shear source function also finds contributions at larger $k$ values from all lower redshifts $z<z_i$ where $\chi(z) < \chi(z_i)$.

We therefore expect the shear angular power spectrum to have more constraining power for $\mathrm{d} n_\mathrm{s}/\mathrm{d}\ln k$ simply because of the longer lever arm. Later we shall see that this is indeed the case.

\subsection{Cluster mass function}\label{sec:cl}

As our cluster observable we use the number of clusters $N_{i,j(i)}$ in the redshift bin $i$ and the mass bin $j$ (depending on $i$), defined as,
\begin{equation}
N_{i,j(i)} = \Delta\Omega\int_0^\infty   \mathrm{d}z \int_0^\infty \mathrm{d}M \
\frac{\mathrm{d}^2V}{\mathrm{d}\Omega \mathrm{d}z}\left(z\right)W_{i,j(i)} (M,z)\frac{\mathrm{d}n_{\rm ST}}{\mathrm{d}M} (M,z),
 \label{eq:Nalbe}
\end{equation}
where $\Delta\Omega$ is the solid angle covered by the survey, $\mathrm{d}^2V/\left(\mathrm{d}\Omega\mathrm{d}z\right)\left(z\right)$ the comoving volume element at redshift $z$, $W_{i,j(i)}\left(M,z\right)$ is the window function defining the redshift (index $i$) and redshift-dependent mass bin (index $j(i)$), and 
\begin{equation}
	\frac{\mathrm{d}n_{\rm ST}}{\mathrm{d}M}(M,z)=\sqrt{\frac{1}{2\pi}}A\left[1+\left(a\nu\right)^{-p}\right]\frac{\bar\rho_\mathrm{m}}{M^2}\sqrt{a\nu}\frac{\mathrm{d}\log \nu}{\mathrm{d}\log M}\exp\left[-a\frac{\nu}{2}\right],
 \label{eq:sh}
\end{equation}
is the Sheth--Tormen mass function~\cite{Sheth:1999mn}.  Here, $a=0.707$, $A=0.322184$, and $p=0.3$, while 
\begin{equation}
\nu\left(M,z\right) \equiv \delta_c^2\left(M,z\right)/\sigma_m^2\left(M,z\right),
\end{equation}
where $\delta_c\left(M,z\right)$ is the linear threshold density established from the spherical collapse model (see~\cite{Basse:2013zua} for details), and
\begin{equation}
\label{eq:sigma}
\sigma_m^2 (M,z) =\int_0^\infty \mathrm{d}\ln k\;\left|W(k R)\right|^2 \Delta^2\left(k,z\right)
\end{equation}
is the variance of the linear matter density field  smoothed with a real-space top-hat filter on a length scale $R = [3 M/(4\pi \bar{\rho}_m)]^{1/3}$ corresponding to the cluster mass~$M$.
The function $W(kR)$ is the Fourier transform of the real-space top-hat filter, and $\Delta^2(k,z)= k^3 P_\delta(k,z)/(2 \pi^2)$ is the linear dimensionless matter power spectrum.


\subsubsection{Measurement errors}

Photometric cluster observations are subject in principle to errors in both the determination of the cluster redshift and the cluster mass. 
However, spectroscopic follow-up on the detected clusters is expected to reduce the redshift uncertainty to a negligible fraction of the redshift bin widths employed in this study.
We therefore treat the redshift as infinitely well-determined.

\paragraph{Weak lensing mass determination}

A Euclid-like survey will detect clusters through their lensing signal.  The minimum cluster mass that can be detected, or the mass detection threshold~$M_\mathrm{thr}\left(z\right)$, is a function of redshift, and can be estimated 
following the procedure of~\cite{Wang:2004pk,Hamana:2003ts} outlined in~\cite{Basse:2013zua}.

In addition, the weak lensing inferred mass of a cluster $M_\mathrm{obs}$ is subject to scatter and bias with respect to the true cluster mass $M$. We assume that the bias can be controlled to the required level of accuracy and model the scatter in the observed mass with a log-normal distribution according to~\cite{Bahe:2011cb,Becker:2010xj},
\begin{equation}
P\left(M_{\rm obs}|M\right) = \frac{1}{M_{\rm obs} \sqrt{2\pi\sigma^2}}\exp\left[-\frac{\left(\ln M_{\rm obs}-\mu\right)^2}{2\sigma^2}\right],
\label{eq:log-normal}
\end{equation}
where $P\left(M_{\rm obs}|M\right)$ is the probability of inferring mass $M_\mathrm{obs}$ under the condition that the true cluster mass is $M$, 
and $\mu = \ln M -\sigma^2/2$ so that the mean of the distribution corresponds to the true mass~$M$.
 The mass scatter $\sigma$ is determined by $N$-body simulations~\cite{Bahe:2011cb}. Integrating this distribution over $M_\mathrm{obs}$ in the observed mass interval $[M_{{\rm min},j(i)}, M_{{\rm max},j(i)}]$
of the $j$th mass bin and accounting  for the mass detection threshold $M_{\rm thr}(z)$ and redshift binning, 
the window function $W_{i,j(i)} (M,z)$ for the cluster survey can be constructed as
\begin{equation}
W_{i,j(i)} (M,z) = \theta(z-z_{\min,i}) \theta(z_{\max, i}-z)   \int_{M_{\min,j(i)}}^{M_{\max,j(i)}} \mathrm{d} M_{\rm obs} \
P\left(M_{\rm obs}|M\right) \theta(M_{\rm obs}-M_{\rm thr}(z)),
\label{eq:probinbin}
\end{equation}
where $\theta$ is the Heaviside step function. 

\paragraph{Completeness and efficiency}

Cluster survey also suffer from incomplete detection of clusters and false cluster detections. 
The errors associated with these effects are characterised respectively by 
the completeness~$f_\mathrm{c}$,  defined as the fraction of clusters actually detected as peaks by the cluster finding algorithm, 
and the efficiency~$f_\mathrm{e}$, defined as the fraction of detected peaks that correspond to real clusters.   In general these quantities can be established precisely only with the help of mock cluster catalogues generated from $N$-body simulations (see, e.g.,~\cite{White:2001gs,Feroz:2008rm,Pace:2007ac,Hennawi:2004ai,Lima:2005tt,Hamana:2003ts}).
Here, we follow the simplistic approach of~\cite{Wang:2005vr}, and assume both $f_\mathrm{c}$ and $f_\mathrm{e}$ to be mass- and redshift-independent. 
Then, both $f_\mathrm{c}$ and $f_{\rm e}$ can be folded into the likelihood function in a simple way (see section~\ref{sec:likelihood}).


\subsubsection{Effective $k$-range of the cluster data}

As in section~\ref{sec:effectivek}, we can ask here what is the effective $k$-range probed by the cluster mass function~(\ref{eq:sh}).  At first glance 
$dn_{\rm ST}/dM$ does not appear to depend on $k$.  However, upon closer inspection, we see that the variance $\sigma_m^2(M,z)$ appearing 
in the variable~$\nu$ is in fact constructed from integrating the linear dimensionless matter power spectrum  $\Delta^2(k,z)$ filtered by a function~$W(kR)$ in equation~(\ref{eq:sigma}) for 
each value of~$M$.  The combination~$|W(kR)|^2 \Delta^2(k,z)/\mathcal{P}_\mathcal{R}(k)$, then, is the counterpart of the source functions we examined previously in section~\ref{sec:effectivek}.

\begin{figure}[t]
\center
\includegraphics[height=.66\textwidth,angle=0]{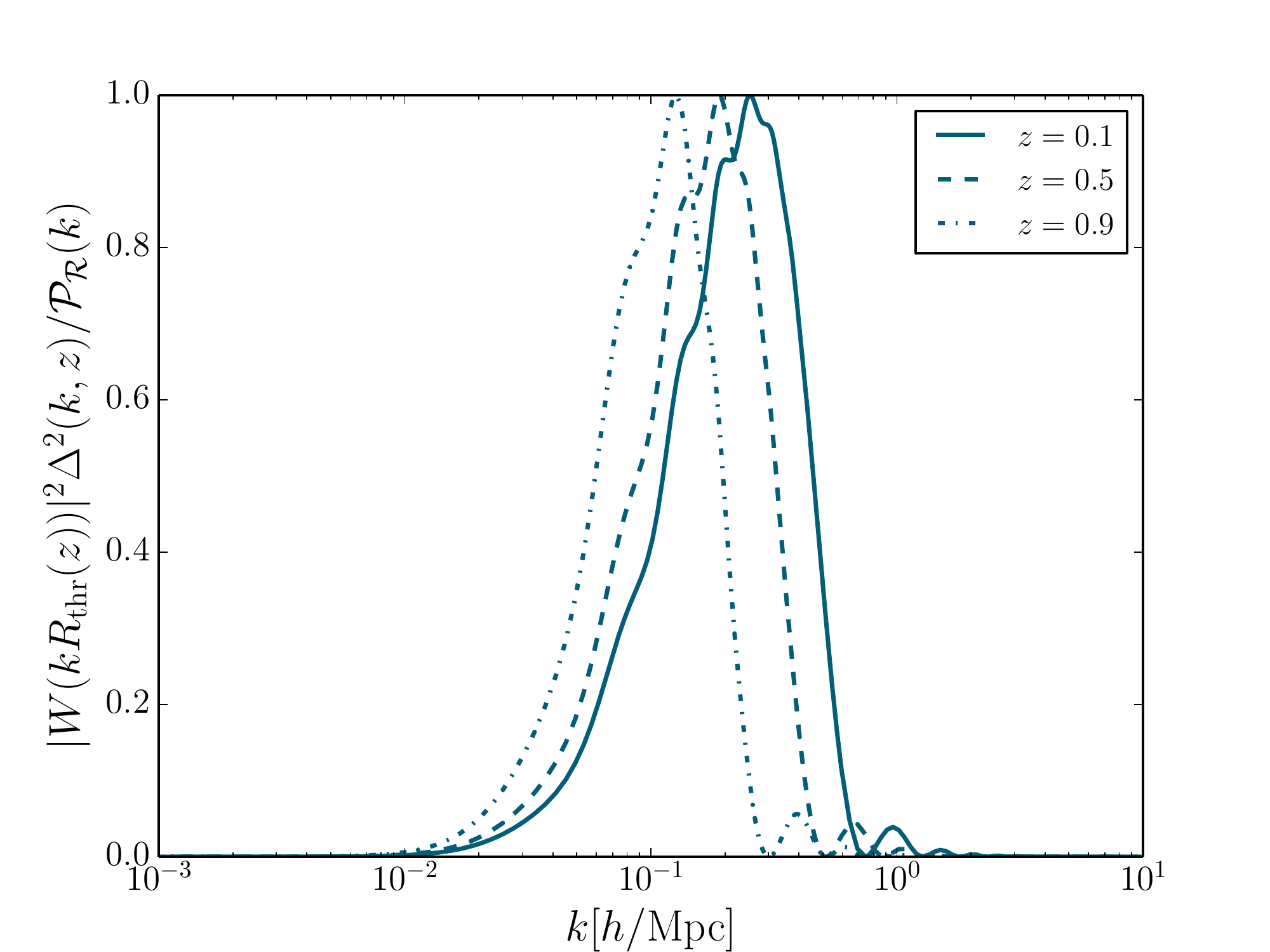}
\caption{The filter function~$|W(kR)|^2 \Delta^2(k,z)/\mathcal{P}_\mathcal{R}(k)$ for three redshift bins, as functions of $k$.  The filter scale $R$ has been chosen to correspond to the cluster mass detection threshold  at the redshift concerned.
 All curves have been normalised to a peak value of unity.~\label{fig:Clsource}}
\end{figure}

Figure~\ref{fig:Clsource} shows $|W(kR)|^2 \Delta^2(k,z)/\mathcal{P}_\mathcal{R}(k)$ at the cluster mass detection threshold in three redshift bins.  Defined this way, we see that the maximum $k$-reach of a
cluster survey is comparable to that of galaxy clustering.

\subsection{CMB temperature and polarisation power spectra}

In order to realistically assess future parameter sensitivities, we must consider CMB an-isotropy measurements in addition to the 
 large-scale structure observables discussed above.  Indeed, some cosmological parameters, notably the tensor-to-scalar ratio~$r$, are accessible almost exclusively to CMB measurements alone. 
 
Information from the CMB can, to some extent, be introduced into a parameter sensitivity forecast in the form of external priors on the relevant cosmological parameters.   However, we caution against this shortcut for the following reasons.
Firstly, such an approach can only make sense if the external priors have been derived under the same model assumptions as the cosmological model we wish to analyse.  
Such priors may be easy to find in the literature for simple, standard cosmological models, but are rarely available for more complicated parameter spaces.
Secondly, even if the first condition can be satisfied, 
 the marginalisation and interval construction procedures employed in the construction of a conventional one-dimensional Gaussian prior
  remove important information about the shape of the likelihood function and the correlations between parameters.  Highly non-Gaussian likelihoods and/or significant correlations between parameters can strongly impact on the outcome of a parameter sensitivity forecast.

The fundamental observable we work with are the CMB power spectra $C^{xy}_\ell$, where $x,y=T, E,B$, representing, respectively, temperature, $E$-polarisation, and $B$-polarisation.  Parity conservation implies $C^{TB}_\ell=C^{EB}_\ell=0$.  We will therefore only work with the three auto-correlation power spectra  $C^{TT}_\ell, C^{EE}_\ell$
and  $C^{BB}_\ell$, and the temperature--$E$-polarisation cross-correlation power spectrum.
The use of synthetic CMB data for parameter sensitivity forecast has been discussed extensively in~\cite{Perotto:2006rj}.   We follow the same approach in this work.


\subsection{Synthetic data sets}\label{sec:datasets}

Having gone through the different types of data we briefly summarise how we construct mock data for each type of observable as well as their possible cross-correlations.

\begin{itemize}

\item The cosmic shear auto-spectrum is $C_{\ell,ij}^{\rm ss}$, with $2 \leq \ell \leq \ell_{\rm max}^{\rm s}=2000$, independently of redshift. As discussed in~\cite{Hamann:2012fe}, shot-noise always dominates well before $\ell$ reaches $2000$, so that the value of the actual cut-off is immaterial.
The indices $i,j \in [1,N_{\rm s}]$ label the redshift bin, where the redshift slicing is such that all bins contain similar numbers of source galaxies and so suffer the same amount of shot noise.  In \cite{Hamann:2012fe} it was found that, for most cosmological parameters, slicing the shear data beyond two redshift bins offers no significant improvement on the sensitivities.  We shall test this assumption for the inflationary parameters studied here.

\item The galaxy auto-spectrum $C_{\ell,ij}^{\rm gg}$ uses multipole moments running from $\ell=2$ to $\ell^{{\rm g},i}_{\rm max}$ in redshift bins $i,j \in [1,N_{\rm g}]$, where $\ell^{{\rm g},i}_{\rm max}$ is set following the prescription of~\cite{Hamann:2012fe} so as to keep the non-linear contamination to a minimum.
It was found in~\cite{Hamann:2012fe} that most cosmological parameter sensitivities saturate at around $N_{\rm g}=11$. Again, we shall test whether or not this limit holds true also for inflationary parameters.   

Regarding the galaxy bias, a realistic bias model will likely be a continuous and monotonic function of redshift, probably described by two or three parameters, calibrated, within errors, against simulations.  In the absence of a concrete model, however, we opt to work in the optimistic extreme and assume the galaxy bias in each redshift bin to be  perfectly known.  
The opposite extreme would be to assume the bias to be unknown and uncorrelated between redshift bins.  However, this equally unrealistic assumption would lead to  parameter constraints that are completely uncompetitive if the galaxy data were used on their own, and, if analysed in combination with shear data, hardly better than the shear-only constraints~\cite{Hamann:2012fe}.  We therefore do not consider  the case of an unknown galaxy bias.

\item The shear--galaxy cross-spectrum $C_{\ell,ij}^{\rm sg}$ in the shear redshift bin $i \in [1,N_{\rm s}]$ and galaxy redshift bin $j \in [1,N_{\rm g}]$ 
runs from $\ell=2$ to $\ell^{{\rm g},j}_{\rm max}$, where  $\ell^{{\rm g},j}_{\rm max}$ is determined by the galaxy redshift binning.

\item The cluster data set~$N_{i,j(i)}$ contains clusters in the redshift range $z \in [0.01, z_{\rm high}]$ and with 
 masses in the range \mbox{$M \in [ M_{\rm thr}(z), 10^{16}~M_\odot]$},
where $M_{\rm thr}(z)$ denotes the redshift-dependent mass detection threshold, and $z_{\rm high}$ is defined as the redshift at which
$M_{\rm thr}(z)$ exceeds $10^{16}~M_\odot$.   Following~\cite{Basse:2013zua}, the redshift bins~$i \in [1,N_z]$ are divided such that for the fiducial model all redshift bins contain equal numbers of clusters.  Clusters in each redshift bin~$i$ are further subdivided into mass bins~$j(i) \in [1,N_\mathrm{m}]$, where, again, the bin boundaries are chosen so that the same number of clusters falls into each bin~\cite{Basse:2013zua}. 
We shall determine in this work how many bins are required in order to extract the maximum possible information from the cluster data.

\item A mock data set from a {\sc Planck}-like CMB measurement is generated according to the procedure of~\cite{Perotto:2006rj}.  This simplistic framework obviously does not allow us to reproduce the full complexity of the published Planck temperature likelihood~\cite{Ade:2013kta}.  Nonetheless, it can be adapted so as to capture some of the gross features of the Planck likelihood, such as the fact that its effective sky coverage is scale-dependent.  We outline below our adaptation and justification.

 The Planck likelihood consists of a low-multipole part ($\ell < 50$) using information from 91\% of the sky (but includes additional component separation uncertainties), and a 
 high-multipole part, which has a sky coverage of 58\% in the 100~GHz and 37\% in the 143~GHz and 217~GHz channels.  Up to around $\ell \sim 1000$, the 100~GHz channel is limited only by cosmic variance and therefore dominates the information.  On smaller scales, the 100~GHz channel becomes noise-dominated;  information on these scales 
therefore comes mainly from the 143~GHz and 217~GHz channels and the effective sky coverage drops to 37\%.

We approximate these behaviours by generating our mock data in two components:
one component running in the range $2 \leq \ell \leq 1000$ with $f_\mathrm{sky} = 0.58$ and the noise properties of the 100~GHz channel (see table~\ref{tab:planckspecs}), and a second in the range $1001 \leq \ell \leq \ell_\mathrm{max}$ with $f_\mathrm{sky} = 0.37$ and the noise properties of the 143~GHz and 217~GHz channels.
We generate mock power spectra for $T$, $E$- and $B$-polarisation auto-correlation, as well as $TE$ cross-correlation, assuming that temperature and polarisation are measured on identical parts of the sky.

The primary CMB signal becomes swamped by foreground anisotropies on the smallest scales.
Rather than modelling explicitly these foregrounds, we opt to mimic their effects on parameter estimation by 
introducing a small scale cutoff, $\ell_\mathrm{max}$.  Using simulated temperature data plus a Gaussian prior on $\tau$, we find that we can recover the parameter sensitivities of real Planck data with $\ell_\mathrm{max} = 1400$.

\begin{table*}[t]
  \caption{Experimental specifications assumed in our Planck data simulations.  The beam width $\theta_\mathrm{FWHM}$ and temperature pixel noise $\Delta_T$ correspond to Planck's actual in-flight performance~\cite{Ade:2013ktc}, while the polarisation pixel noise $\Delta_P$ is set to the design performance goal given in~\cite{Planck:2006aa}. \label{tab:planckspecs}}
\begin{center}
{\footnotesize
  \hspace*{0.0cm}\begin{tabular}
  {cccc} \hline \hline
  Frequency/GHz & $\theta_\mathrm{FWHM}/\mathrm{arcmin}$ & $\Delta_T/\mu\mathrm{K}$ & $\Delta_P/\mu\mathrm{K}$ \\ \hline 
  100 & 9.59 & 11.0 & 10.9\\
  143 & 7.18 & 6.0 & 11.4\\
  217 & 4.87 & 12.0 & 26.7\\
  \hline \hline
  \end{tabular}
  }
  \end{center}
\end{table*}

\end{itemize}


\section{Forecasting \label{sec:forecast}}

We now describe our parameter sensitivity forecast for a  Euclid-like photometric survey including a measurement of the  cluster mass function.
The forecast is based on the construction of a likelihood function for the mock data, whereby the survey's sensitivities to cosmological parameters can be explored using
Bayesian inference techniques.


\subsection{Model parameter space\label{sec:fiducial}}

The forecast code has previously been used to probe extensions of the $\Lambda$CDM model, either in the neutrino sector \cite{Hamann:2012fe}
or in dark energy \cite{Basse:2013zua}.
Here, we take a similar approach, and extend the standard $\Lambda$CDM model in the inflation sector.
The non-inflationary parameters of the extended model are the physical baryon density $\omega_\mathrm{b}$, the physical cold dark matter  density $\omega_{\rm c}$, the dimensionless Hubble parameter $h$, and the reionisation redshift  $z_{\rm re}$; the inflation sector of the model is  described by the amplitude of the scalar fluctuation spectrum~$A_\mathrm{s}$, the spectral index~$n_\mathrm{s}$, the running of the spectral index~$\mathrm{d}n_\mathrm{s}/\mathrm{d}\ln k$, and the tensor-to-scalar ratio~$r$, all defined at the pivot scale.  Thus, in total the model has eight free parameters:
\begin{equation}
\label{eq:model}
\Theta^{(8)} \equiv \Theta^{(\rm non-infl)} + \Theta^{(\rm infl)} 
 \equiv \left( \omega_{\rm b},\omega_{\rm c},h,z_{\rm re}, A_\mathrm{s}, n_\mathrm{s}, \mathrm{d}n_\mathrm{s}/\mathrm{d}\ln k, r \right).
\end{equation}
For the non-inflationary part of the parameter space, our fiducial model is defined by  the parameter values
\begin{equation}
\Theta^{({\rm non-infl})}_{\rm fid} = (0.0226, 0.112, 0.7, 11).
\end{equation}
For the inflationary parameter space we will study a variety of fiducial models, but
\begin{equation}
\label{eq:fiducial}
\Theta^{({\rm infl})}_{\rm fid} = (2.1 \times 10^{-9}, 0.96, 0, 0.1)
\end{equation}
will be  our most common choice.


\subsection{Likelihood functions}\label{sec:likelihood}

\paragraph{CMB}

The effective $\chi^2$ from CMB can be constructed using the method outlined in~\cite{Perotto:2006rj}, and is given by
\begin{equation}
\chi^2_{\rm eff} = \sum_{\ell} (2\ell+1) f_{\rm sky} \left[{\rm Tr}(\bar{\mathbf C}_{\ell}^{-1} \hat{\mathbf C}_{\ell}) + \ln \frac{{\rm Det}(\bar{\mathbf C}_{\ell})}{{\rm Det} (\hat{\mathbf C}_{\ell})} - N \right],
\label{eq:chi2cmb}
\end{equation}
where $\hat{\mathbf{C}}_\ell \doteq \hat{C}_\ell^{xy}$ is the $(N\times N)$-dimensional mock data covariance matrix, $\bar{\mathbf C}_\ell \doteq C_\ell^{xy}$ the model covariance matrix, and the expression has been normalised such that $\chi_{\rm eff}^2=0$ when  $\hat{\mathbf C}_\ell=\bar{\mathbf{C}}_\ell$. 
We set $N=2$ when only $T$ and $E$ data are used, and 3 when $B$ data are also included.%
\footnote{Our modelling of the CMB likelihood function assumes the signal to be Gaussian-distributed, and clearly does not capture the large non-Gaussian covariance of the lensed $B$-mode~\cite{Smith:2004up}.  This non-Gaussianity is very important for future high-resolution, low-noise $B$-polarisation experiments~\cite{Li:2006pu}, but can be neglected for experiments of Planck-like sensitivity, such as considered in this work. }
In all cases we work with lensed power spectra.

\paragraph{Shear and galaxies}

The effective $\chi^2$ for shear and galaxies has exactly the same form as equation~(\ref{eq:chi2cmb}), but with $\hat{\mathbf C}_\ell$ and $\bar{\mathbf C}_\ell$ identified with $\hat{C}^{XY}_{\ell, ij}$ and $\bar{C}^{XY}_{\ell, ij}$ respectively, where $X,Y={\rm s}, {\rm g}$.  The dimension parameter $N$ is determined by the number of redshift bins used in the analysis:  $N=N_{\rm s}$ for shear-only (s), $N_{\rm g}$ for galaxies-only (g), and $N_{\rm s}+N_{\rm s}$ if we are interested in shear--galaxy cross-correlation (x) as well.

\paragraph{Clusters}

The cluster likelihood function in one redshift and mass bin can be modelled at the most basic level as a Poisson distribution in the observed number of clusters~$N_{\rm obs}$, with the theoretical prediction $N_{\rm th}$ as the mean.
However, the imperfect completeness and efficiency of the survey necessitate that we rescale the uncertainty on each data point by an amount $f^{-1} \equiv \sqrt{[1/f_\mathrm{e}+(1/f_\mathrm{e}-1)]/f_\mathrm{c}}$ (see \cite{Basse:2013zua}).  We accomplish this by defining an effective number of observed clusters \mbox{$\widetilde{N}_{\rm obs} \equiv f^2 N_{\rm obs}$}, and likewise an effective theoretical prediction \mbox{$\widetilde{N}_{\rm th} \equiv f^2 N_{\rm th}$}.  Then, the effective probability distribution is
\begin{equation}
\label{eq:poison}
\mathcal{L}_\mathrm{P}\left(\widetilde{N}_\mathrm{obs}|\widetilde{N}_\mathrm{th}\right) = \frac{\widetilde{N}_\mathrm{th}^{\widetilde{N}_\mathrm{obs}}}{\widetilde{N}_\mathrm{obs}!} \, \exp \left[ -\widetilde{N}_\mathrm{th} \right].
\end{equation}
In a real survey, the effective observed number of clusters $\widetilde{N}_\mathrm{obs}$ in any one bin is necessarily an integer so that equation~(\ref{eq:poison}) applies directly.  In our forecast, however, $\widetilde{N}_\mathrm{obs}$ corresponds to the theoretical expectation value of the fiducial model which generally does not evaluate to an integer.  To circumvent this inconvenience, we generalise  the likelihood function~(\ref{eq:poison}) by linearly interpolating the logarithm of the discrete distribution $\mathcal{L}_\mathrm{P}$
 in the interval $[ \mathrm{floor} ( \widetilde{N}_\mathrm{obs}),  \mathrm{ceiling} ( \widetilde{N}_\mathrm{obs}) ]$, i.e.,
\begin{equation}
\begin{aligned}
\ln {\mathcal{L}}( \widetilde{N}_\mathrm{obs}|\widetilde{N}_\mathrm{th})
 \equiv & \left(1  + \mathrm{floor} (\widetilde{N}_\mathrm{obs}) - \widetilde{N}_\mathrm{obs} \right)  \ln \mathcal{L}_\mathrm{P}\left(\mathrm{floor} (\widetilde{N}_\mathrm{obs} )|\widetilde{N}_\mathrm{th} \right)  \\
& + \left( \widetilde{N}_\mathrm{obs} - \mathrm{floor} ( \widetilde{N}_\mathrm{obs} ) \right) \ln \mathcal{L}_\mathrm{P} \left( \mathrm{ceiling} (\widetilde{N}_\mathrm{obs})|\widetilde{N}_\mathrm{th} \right).
\end{aligned}
\end{equation}
The total cluster log-likelihood function is then obtained straightforwardly by summing $\ln {\mathcal{L}}$ over all redshift and mass bins.


\section{Results}
\label{sec:results}

\subsection{Binning}

Before proceeding with a more detailed discussion of our results, we first test how our choice of redshift and mass binning affects parameter constraints on $n_\mathrm{s}$ and $\mathrm{d} n_\mathrm{s}/\mathrm{d}\ln k$ for each of the three large-scale structure observables. The study is performed in the same manner as in \cite{Hamann:2012fe} and \cite{Basse:2013zua}, i.e., we combine each of the three data sets with CMB data and test how parameter constraints vary with the number of reshift (shear, galaxy, and cluster data) and mass (cluster data) bins.

\begin{figure}[t]
\center
\includegraphics[height=.66\textwidth,angle=0]{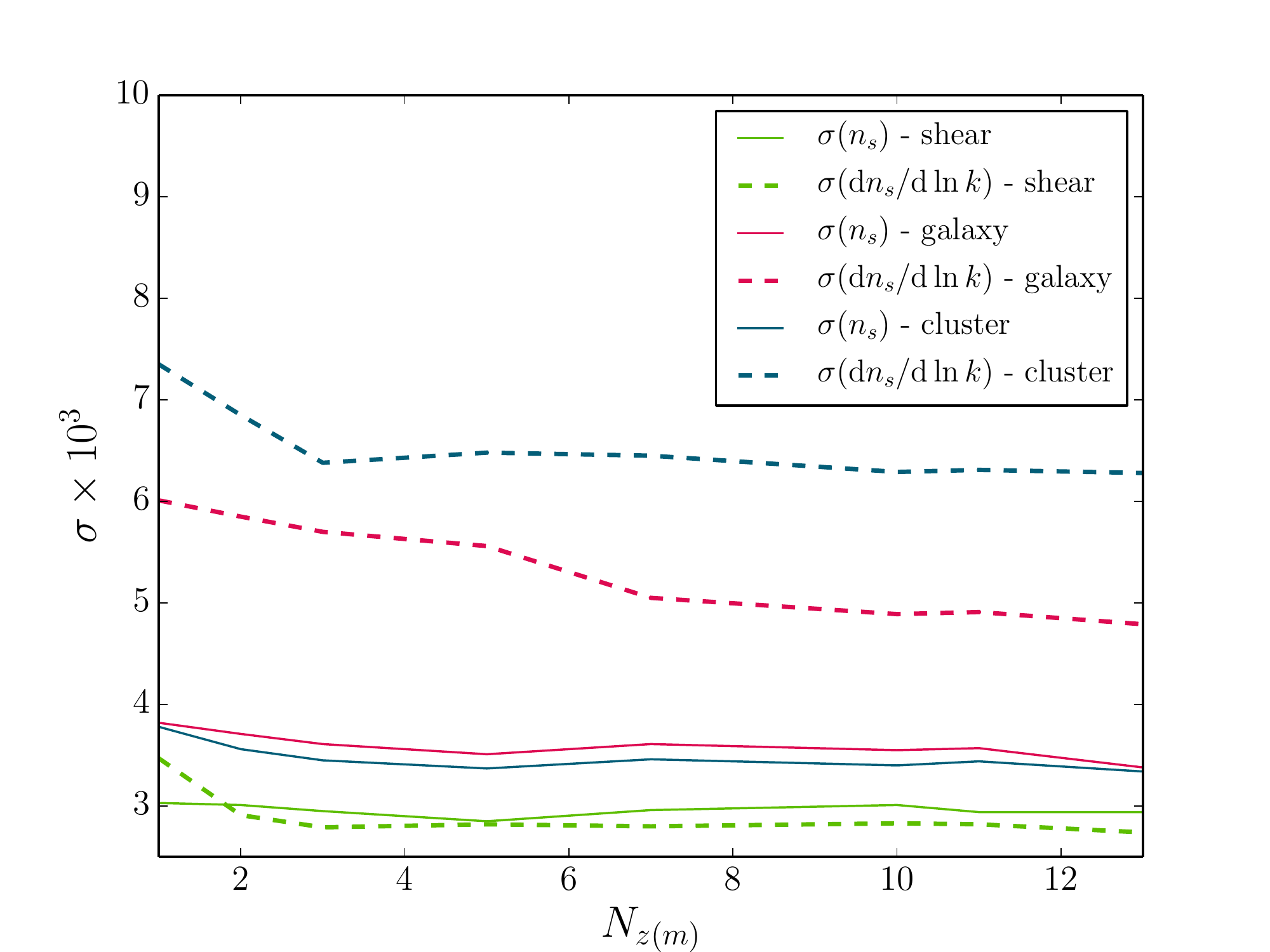}
\caption{ 
Standard deviation of the one-dimensional marginalised posterior distribution for $n_\mathrm{s}$ (solid lines) and for $\mathrm{d} n_\mathrm{s}/\mathrm{d}\ln k$ (dashed lines) as  a function of the number of redshift bins $N_z$, derived from various data combinations. Green curves are for CMB+shear, red for CMB+galaxies, and blue for CMB+clusters. Note that for clusters we fix the number of mass bins to be the same as the number of redshift bins.
\label{fig:bins}}
\end{figure}

Figure~\ref{fig:bins} shows the parameter constraints when varying the number of redshift bins from 1 to 13.  In the case of clusters, we also fix, for simplicity, the number of mass bins $N_{\rm m}$  to be the same as the number of redshift bins~$N_z$.  For an example of how the choice of disparate $N_{\rm m}$ and $N_z$ values affects neutrino and dark energy constraints, see~\cite{Basse:2013zua}.

The results in figure~\ref{fig:bins} confirm the trends found in~\cite{Hamann:2012fe,Basse:2013zua}, namely, that  parameter constraints do not improve significantly beyond 
two redshift bins for shear data, about ten bins for galaxies, and $N_z=N_{\rm m} \sim 2 \to 3$ for clusters.
In fact, because we are considering parameters related to the initial conditions in this study, rather than parameters that affect the late-time evolution of the observables (e.g., the dark energy equation of state), the redshift binning is even less important here than was found in~\cite{Hamann:2012fe,Basse:2013zua}.  For ease of comparison, however, 
we shall from now on use the same binning choices as~\cite{Basse:2013zua}, i.e., $N_{\rm s} = 2$, $N_{\rm g} = 11$, and $N_{z}=N_{\rm m}=10$.
We emphasise again, however, that none of our results depends strongly on the binning choice.


\subsection{Breaking degeneracies \label{sec:deg}}

Results from our main runs are shown in table~\ref{tab:errors}, where we have derived parameter sensitivities from a Planck-like CMB experiment plus the three large-scale structure observables, individually as well as in combination.  Clearly, the sensitivities to $n_{\rm s}$ and $\mathrm{d} n_\mathrm{s}/\mathrm{d}\ln k$ reflect strongly on the 
effective $k$-range available to the observables used: while adding galaxy or cluster data to CMB offer some improvement over the CMB-only sensitivities, both are out-performed by CMB+shear, which,
as shown in figure~\ref{fig:source}, has a much longer lever arm.

\begin{table*}[t]
  \caption{One-dimensional posterior standard deviations for the parameters $A_\mathrm{s}$, $n_\mathrm{s}$, $\mathrm{d} n_\mathrm{s}/\mathrm{d}\ln k$, and~$r$, derived from various data combinations assuming the fiducial model~(\ref{eq:fiducial}).
   Here, ``c'' denotes Planck-like CMB data, ``g'' galaxy auto-correlation (11 redshift bins), ``s'' shear auto-correlation (2 bins), ``x'' shear--galaxy cross-correlation, and ``cl'' cluster data (10 redshift bins, 10 mass bins). }
\label{tab:errors}
\begin{center}
{\footnotesize
  \hspace*{0.0cm}\begin{tabular}
  {lcccc} \hline \hline
  Data & $\sigma(\log A_\mathrm{s})$ & $\sigma(n_\mathrm{s})$ & $\sigma(\mathrm{d} n_\mathrm{s}/\mathrm{d}\ln k)$ & $\sigma(r)$  \\       \hline
  c & 0.011 & 0.0052 & 0.0074 & 0.028\\
  cs & 0.0091 & 0.0030 & 0.0030 & 0.027 \\
  cg & 0.0046 & 0.0035 & 0.0048 & 0.027 \\
  ccl & 0.0068 & 0.0034 & 0.0064 & 0.026 \\
  cscl & 0.0066 & 0.0028 & 0.0029 & 0.026 \\
  csgxcl & 0.0032 & 0.0025 & 0.0017 & 0.026 \\
  \hline \hline
  \end{tabular}
  }
  \end{center}
\end{table*}

\begin{figure}[t]
\center
\includegraphics[height=.35\textwidth]{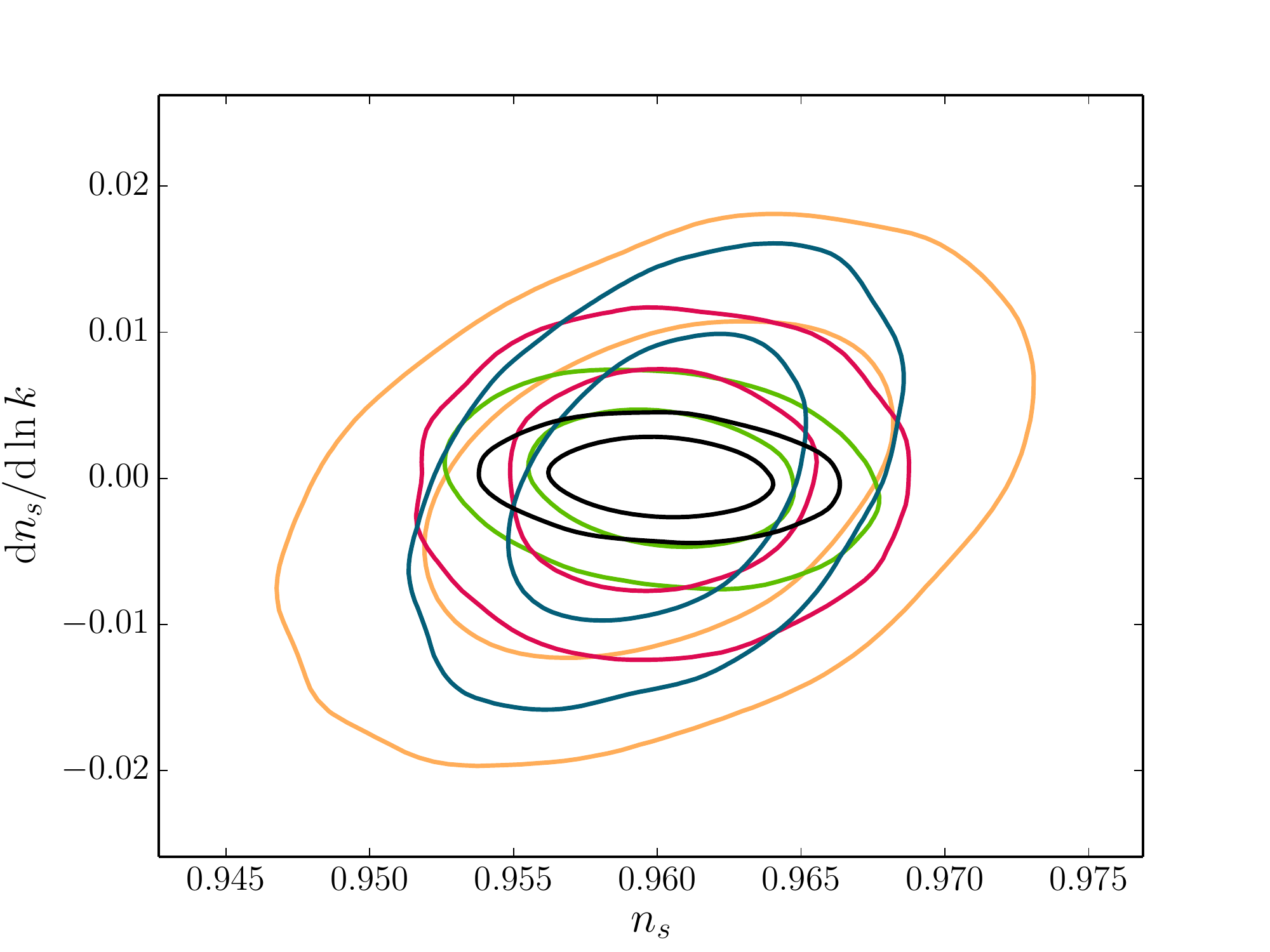}
\includegraphics[height=.35\textwidth]{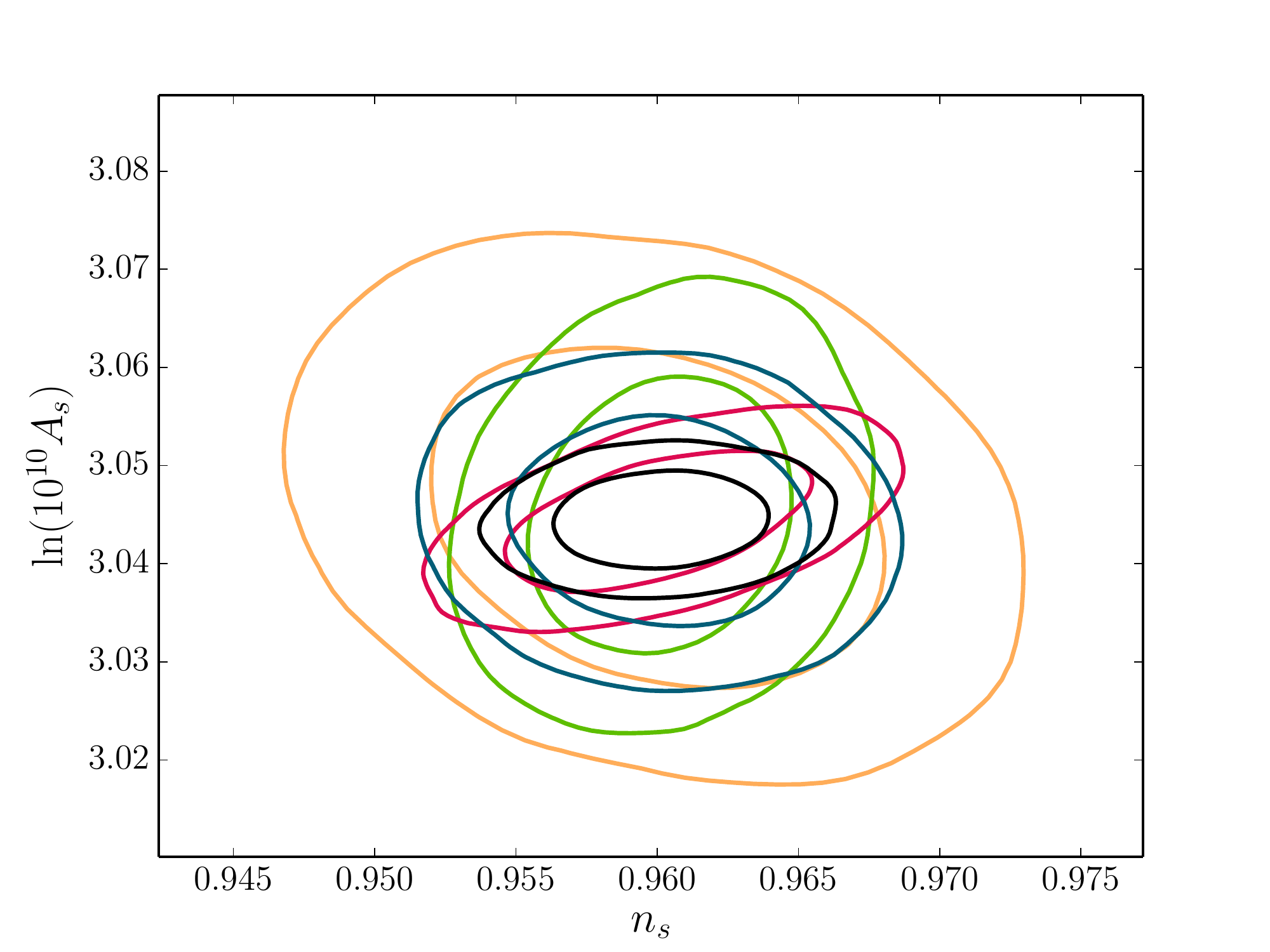} \\
\includegraphics[height=.35\textwidth]{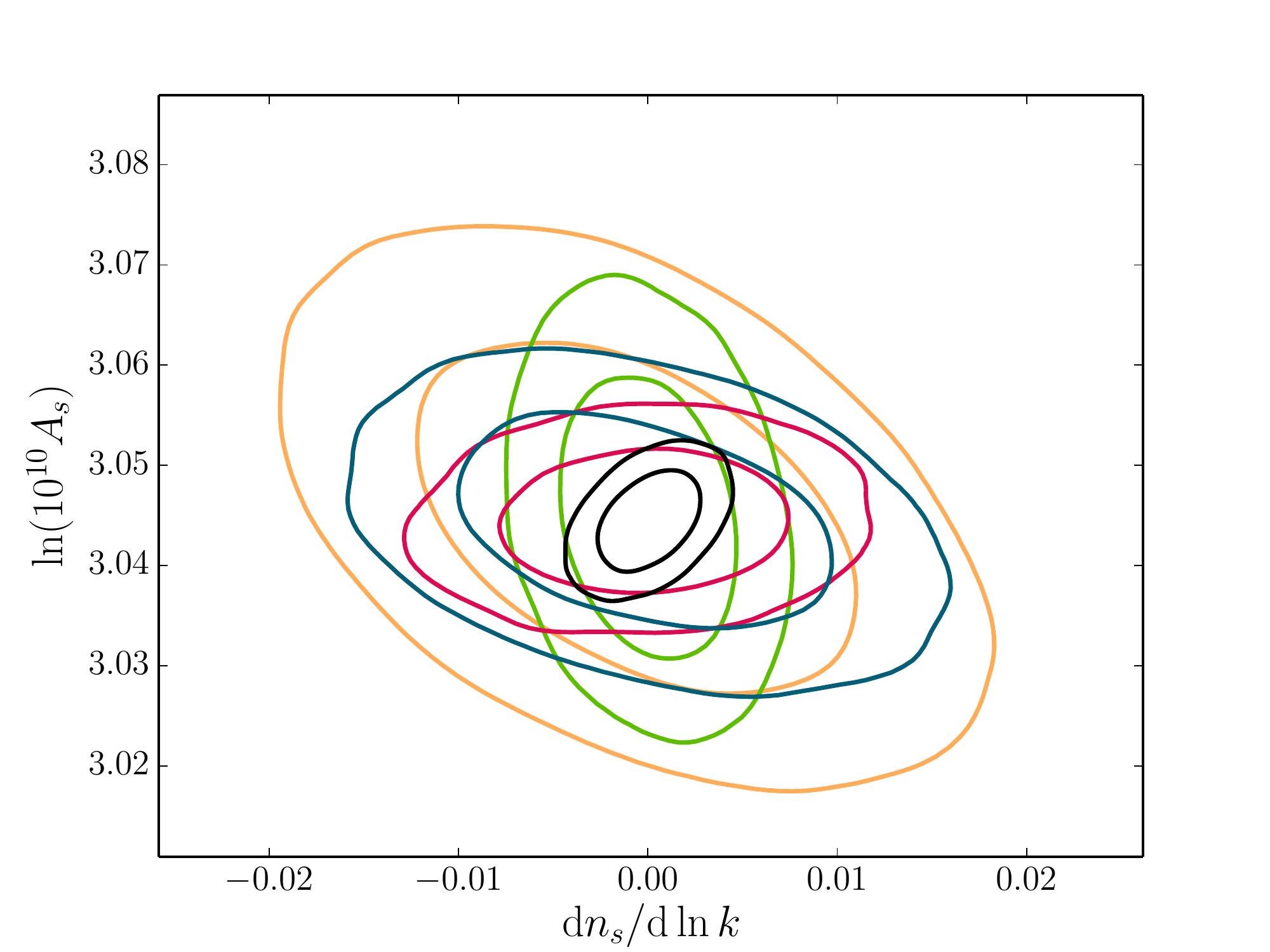}
\includegraphics[height=.35\textwidth]{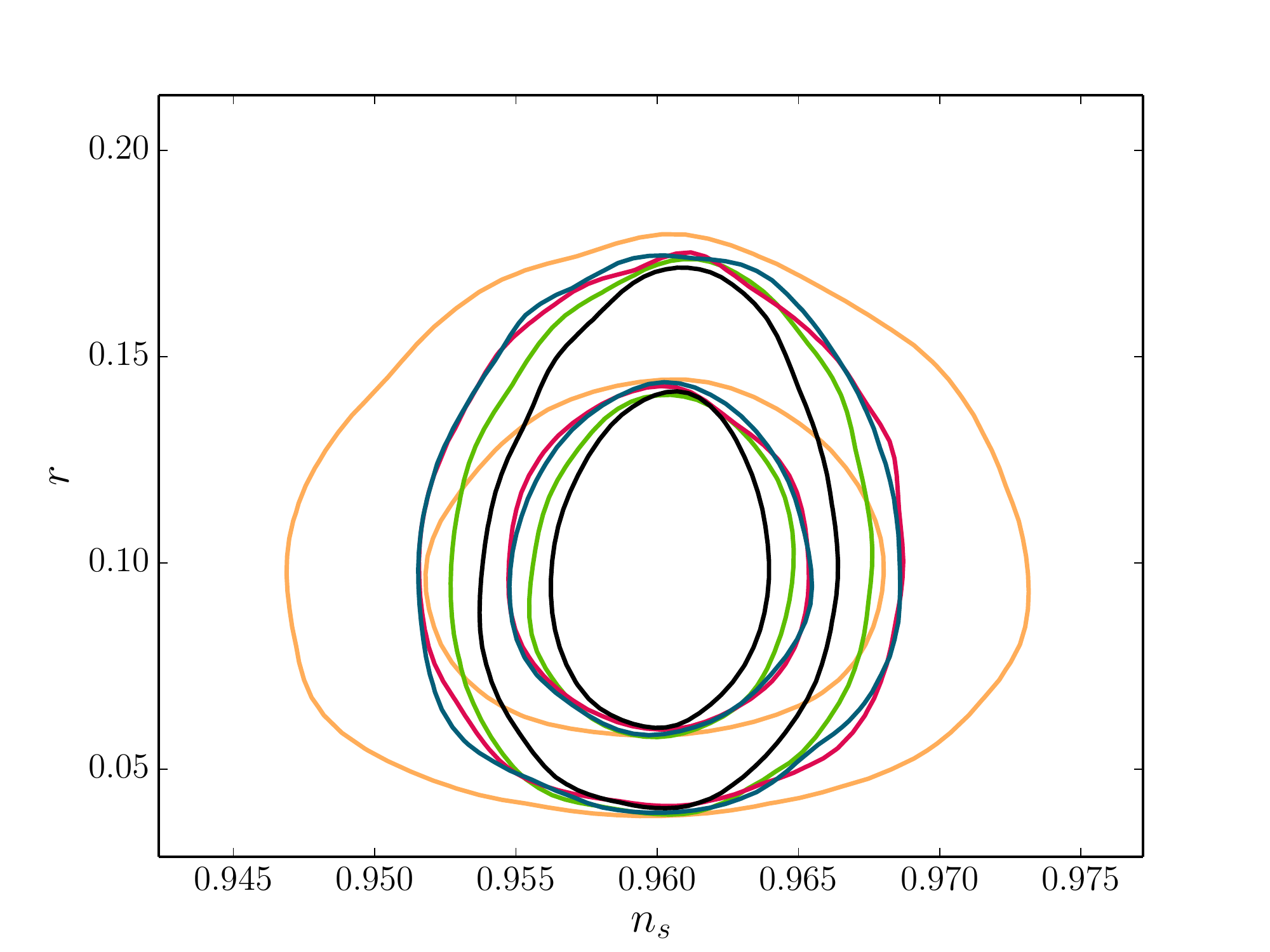}
\caption{Two-dimensional joint marginalised 68\%- and 95\%-credible contours for the inflationary parameters $A_\mathrm{s}, n_\mathrm{s}, \mathrm{d} n_\mathrm{s}/\mathrm{d}\ln k$, and $r$, derived from various data combinations. Yellow lines denote ``c'', red lines ``cs'', blue ``cg'', green ``ccl'', and black ``csgxcl''.\label{fig:degeneracies}}
\end{figure}

Interestingly, even though the sensitivities of galaxy and cluster data {\it per se} are generally inferior to what can be gained from shear measurements, when all used in combination the former two probes in fact contribute to breaking important parameter degeneracies in the shear power spectrum, leading to significant improvements in the the parameter sensitivities over using shear measurements alone.
This effect has already been observed in~\cite{Hamann:2012fe} and~\cite{Basse:2013zua} in relation to neutrino mass constraints.
Our results in table~\ref{tab:errors} show that the same thing is  happening also  to the inflationary parameters (see also figure~\ref{fig:degeneracies}
for the corresponding two-dimensional posteriors).
For example, the sensitivity to ${\rm d} \ln n_{\rm s} /{\rm d} \ln k$ improves from  $0.0030$ for CMB+shear, to $0.0017$ for CMB+shear+galaxy+clusters+cross-correlation.

A caveat is in place here.  As discussed in section~\ref{sec:datasets}, in deriving the our parameter sensitivities we have assumed the linear galaxy bias to be perfectly known in every redshift bin.
This is likely an overly optimistic assumption.  For this reason, in addition to the ``csgxcl'' data combination, we consider also ``cscl'', i.e., {\it without} galaxy data.  The parameter sensitivities derived from this data combination would correspond roughly to those in the (equally unrealistic) case in which the galaxy bias is completely unknown and uncorrelated between the eleven redshift bins.  Under this condition, we see from table~\ref{tab:errors} that most parameter sensitivities essentially revert back to their CMB+shear values, while $\sigma(A_{\rm s})$ mimics its CMB+cluster counterpart.  Thus, how well the angular galaxy power spectrum breaks the parameter degeneracies in other large-scale structure observables does depend strongly on how well the galaxy bias is understood.  Should a viable bias model become available, the true parameter sensitivities would lie somewhere between the ``csgxcl'' and ``cscl'' extremes.


\subsection{Dependence on the fiducial model}

So far we have assumed the fiducial model~(\ref{eq:fiducial}) to be the true model of the universe.  Here, we consider how the parameter sensitivities might change with respect to this assumption.

Table~\ref{tab:errorsfid} shows the sensitivities of the data combination``csgxcl'' to the inflationary parameters with respect to variations in the 
fiducial values of $\mathrm{d} n_\mathrm{s}/\mathrm{d}\ln k$ and $r$.  Evidently, with the exception of~$r$, the parameter sensitivities are quite independent of our choice of fiducial model.   In contrast, the sensitivity to~$r$ is almost a factor of three better when the fiducial model has $r_{\rm fid}=0$, compared with the choice of $r_{\rm fid}=0.1$.
This difference  can be put down to two effects:
 \begin{enumerate}
\item \label{point1}
 If we assume the posterior $P(r)$ to be a Gaussian with width parameter $\sigma$ centred on the fiducial value, then the fact that the distribution is cut off at $r=0$ alone will cause the standard deviation of $P(r)$ to depend on the fiducial value.  The extreme example is $r_\mathrm{fid}$ = 0, in which case $P(r)$ is a half-Gaussian, and the corresponding standard deviation is $(1-2/\pi)^{1/2} \sigma \sim 0.6 \sigma$.  This is in contrast with the case where $r_\mathrm{fid} \gg \sigma$, for which the standard deviation approaches $\sigma$.  By itself, however, this effect cannot entirely account for the observed difference in the sensitivities. 

\item The cosmic variance contribution to the effective uncertainty of the CMB spectra depends on the chosen $r_{\rm fid}$.  
This is generally a negligible effect for the  $T$ and $E$ spectra (even at low~$\ell$), because contributions from scalar perturbations always dominate over tensor contributions. 
For the $B$-polarisation power spectrum, however, which in our case is sourced only by tensor perturbations, the larger cosmic variance following from a larger $r_{\rm fid}$ value
can have an important effect on the sensitivity to $r$.  

 To test this effect, we repeat the analysis for the fiducial models $(\mathrm{d} n_\mathrm{s}/\mathrm{d}\ln k_{\rm fid},r_{\rm fid}) = (0,0)$ and $(0,0.1)$, but this time using only $T$ and $E$ data in the CMB component.  We find $\sigma(r) = 0.036$ and $0.051$, respectively, a much less dramatic disparity than that between  $0.009$ and $0.026$  we found earlier with the $B$-mode included.   In fact, the ratio between  $\sigma(r) = 0.036$ and $0.051$ is approximately $0.7$, which is very close to the ratio of~$0.6$ expected from the effect discussed above in point~\ref{point1}.  This confirms that the larger $B$-mode error bars accompanying a larger $r_{\rm fid}$ value are indeed the main culprit for the reduced sensitivity to $r$.

\end{enumerate}

\begin{table*}[t]
  \caption{One-dimensional posterior standard deviations for the parameters $A_\mathrm{s}$, $n_\mathrm{s}$, $\mathrm{d} n_\mathrm{s}/\mathrm{d}\ln k$, and~$r$, derived from 
  the ``csgxcl'' data combination assuming various fiducial models. See the caption of table~\ref{tab:errors} for an explanation of the data set abbreviations.}
\label{tab:errorsfid}
\begin{center}
{\footnotesize
  \hspace*{0.0cm}\begin{tabular}
  {lcccccc} \hline \hline
  Data & $\mathrm{d}n_s/\mathrm{d}\ln k_{\rm fid}$ & $r_{\rm fid}$ & $\sigma(\log A_\mathrm{s})$ & $\sigma(n_\mathrm{s})$ & $\sigma(\mathrm{d} n_\mathrm{s}/\mathrm{d}\ln k)$ & $\sigma(r)$  \\       \hline
  csgxcl & 0 & 0 & 0.0031 & 0.0025 & 0.0017 & 0.0090 \\
  csgxcl & -0.008 & 0 & 0.0031 & 0.0025 & 0.0018 & 0.0088 \\
  csgxcl & 0 & 0.1 & 0.0032 & 0.0025 & 0.0017 & 0.026 \\
  csgxcl & -0.008 & 0.1 & 0.0032 & 0.0025 & 0.0018 & 0.026 \\
  \hline \hline
  \end{tabular}
  }
  \end{center}
\end{table*}


\subsection{Comparison with other works \label{sec:comp}}

Reference~\cite{Huang:2012mr} (also quoted in~\cite{Amendola:2012ys})  found that Planck-like CMB data combined with data from a Euclid-like spectroscopic survey could attain a sensitivity of $\sigma(n_\mathrm{s}) = 0.0017$ and $\sigma(\mathrm{d} n_\mathrm{s}/\mathrm{d}\ln k) = 0.003$, comparable to our 0.0025 and 0.0021 from the ``csgxcl'' data combination. There are however some subtle differences between their and our analyses.

Firstly, reference~\cite{Huang:2012mr} made much more optimistic assumptions about their CMB data than we do in our study. 
The sensitivities from Planck-like CMB data alone were assumed to be $\sigma(n_\mathrm{s}) = 0.003$ and $\sigma(\mathrm{d} n_\mathrm{s}/\mathrm{d}\ln k) = 0.005$, compared with our 0.0053 and 0.0075.   
Secondly, the addition of large-scale structure observables in our case improves the constraining power by a factor of 2.12 for $n_\mathrm{s}$ and 3.57 for $\mathrm{d} n_\mathrm{s}/\mathrm{d}\ln k$, whereas for the spectroscopic survey studied in \cite{Huang:2012mr} the corresponding factors are 1.76 and 1.7.

That  photometric data carry more weight than do spectroscope data is most likely due to the former's larger effective survey volume and hence statistical power.
After all, parameters like~$n_{\rm s}$ and $\mathrm{d} n_\mathrm{s}/\mathrm{d}\ln k$ do not produce any redshift-dependent effects, and therefore benefit little from high-accuracy redshift measurements.
Rather, they induce slow variations with~$k$, and are hence better constrained  by broad-band power spectrum measurements with high statistics.


\section{Implications for single field slow roll inflation \label{sec:sloro}}

{\it Prima facie}, the running of the spectral index is simply a phenomenological parameter.  However, as it turns out, the large improvements in sensitivity to $\mathrm{d} n_\mathrm{s}/\mathrm{d}\ln k$ afforded by Euclid-like data can potentially have a profound impact on our understanding of the dynamics of slow-roll inflation.  In the slow-roll picture, the four parameters describing the primordial perturbations $\Theta^{(\mathrm{infl})}$ can be expressed in terms of the Hubble parameter and its derivatives during inflation.  In addition, as discussed for instance in~\cite{Easther:2006tv}, the running of the spectral index is intimately related to the total amount of inflation, with large negative values implying that inflation ends prematurely.  In the following section, we briefly remind the reader of the theoretical background.

\subsection{Slow-roll inflation}

The evolution of the Hubble parameter during inflation is determined by an infinite hierarchy of the Hubble slow-roll parameters~\cite{Liddle:1994dx}:
\begin{equation}
\begin{aligned}
\epsilon &\equiv \frac{m^2_\mathrm{Pl}}{4\pi} \left(\frac{H'}{H}\right)^2,\\
^j \lambda_H &\equiv \left(\frac{m^2_\mathrm{Pl}}{4\pi}\right)^j \frac{\left(H'\right)^{j-1}}{H^j}\frac{\mathrm{d}^{\left(j+1\right)}H}{\mathrm{d}\phi^{\left(j+1\right)}},\quad j \geq 1,
\end{aligned}
\end{equation}
where $'$ denotes a derivative with respect to the scalar field $\phi$, and the usual slow roll parameters are $\eta = {^1\lambda_H}$ and $\xi = {^2\lambda_H}$.  Inflation continues as long as $\epsilon < 1$.

If the hierarchy is truncated at $j = 2$ (which is in fact a very good approximation for the majority of generic single-field models of inflation), the solution for the Hubble parameter as a function of $\phi$ then takes the form~\cite{Liddle:2003py}
\begin{equation}
\label{eq:H}
\frac{H}{H_0}= 1+\sqrt{4\pi \epsilon_0}\frac{\phi}{m_\mathrm{Pl}}+2\pi\eta_0\frac{\phi^2}{m_\mathrm{Pl}^2}+\frac{4\pi^{3/2}}{3\sqrt{\epsilon_0}}\xi_0 \frac{\phi^3}{m_\mathrm{Pl}^3},
\end{equation}
where the subscript ``$0$'' refers to the value of the respective quantity at the moment the mode $k_\mathrm{p}$ leaves the horizon.  Without loss of generality we can set $\phi_0 = 0$. 

Given a set of values for the phenomenological observables $n_\mathrm{s}$, $r$ and $\mathrm{d}n_\mathrm{s}/\mathrm{d}\ln k$, the corresponding slow-roll parameters, 
up to and including $j=2$ and up to second order, can be established via the relations~\cite{Kinney:2002qn,Peiris:2006ug},
\begin{equation} 
\label{eq:phentosloro}
\begin{aligned}
n_\mathrm{s} &= 1+2\eta-4\epsilon-2\left(1+\mathcal{C}\right)\epsilon^2-\frac{1}{2}\left(3-5\mathcal{C}\right)\epsilon\eta+\frac{1}{2}\left(3-\mathcal{C}\right)\xi, \\
r &= 16\epsilon\left(1+2\mathcal{C}\left(\epsilon-\eta\right)\right),\\
\mathrm{d}n_\mathrm{s}/\mathrm{d}\ln k &= -\frac{1}{1-\epsilon}\left(2\xi+8\epsilon^2-10\epsilon\eta+\frac{7\mathcal{C}-9}{2}\epsilon\xi + \frac{3-\mathcal{C}}{2}\xi\eta\right),
\end{aligned}
\end{equation}
where $\mathcal{C} = 4\left(\ln 2+\gamma\right)-5$, and $\gamma$ is the Euler--Mascheroni constant.  Solving these relations for $\epsilon$, $\eta$, and $\xi$ then allows for 
 a determination of the number of $e$-foldings until the end of inflation,~$N_e$, via
\begin{equation}
N_e = \frac{4\pi}{m_\mathrm{Pl}^2}\int_{\phi_\mathrm{e}}^{\phi_0}{\mathrm{d}\phi \frac{H}{H'}},
\end{equation}
where $\phi_\mathrm{e}$ is the field value at the end of inflation, i.e., $\epsilon(\phi_\mathrm{e}) = 1$. 

For successful inflation, one typically requires $40 \lesssim N_e \lesssim 60$, with the exact number depending on the details of the reheating process (see, e.g.,~\cite{Ade:2013uln}).  As a consequence, only a certain region of ($n_\mathrm{s}$, $\mathrm{d}n_\mathrm{s}/\mathrm{d}\ln k$, $r$)-space is actually consistent with our assumptions about inflation.  If observational data should decisively favour parameter combinations outside of this region, then some of the assumptions going into expression~(\ref{eq:H}) must be wrong.  One possibility in that case would be the presence of additional, sizeable higher-order terms from the $^j\lambda_H$ hierarchy in order to describe the evolution of $H$ during inflation;
the mere presence of these higher-order terms would rule out most simple models of single-field slow-roll inflation.  Alternatively, one could invoke multi-field dynamics, particularly if data should prefer the $N_e > 60$ region.  Either way, such a discovery would completely change the current picture of inflation.  Let us now turn to exploring the potential of a future Euclid-like survey in this regard.


\subsection{Slow-roll analysis and results}

\begin{figure}[t]
\center
\includegraphics[height=.8\textwidth]{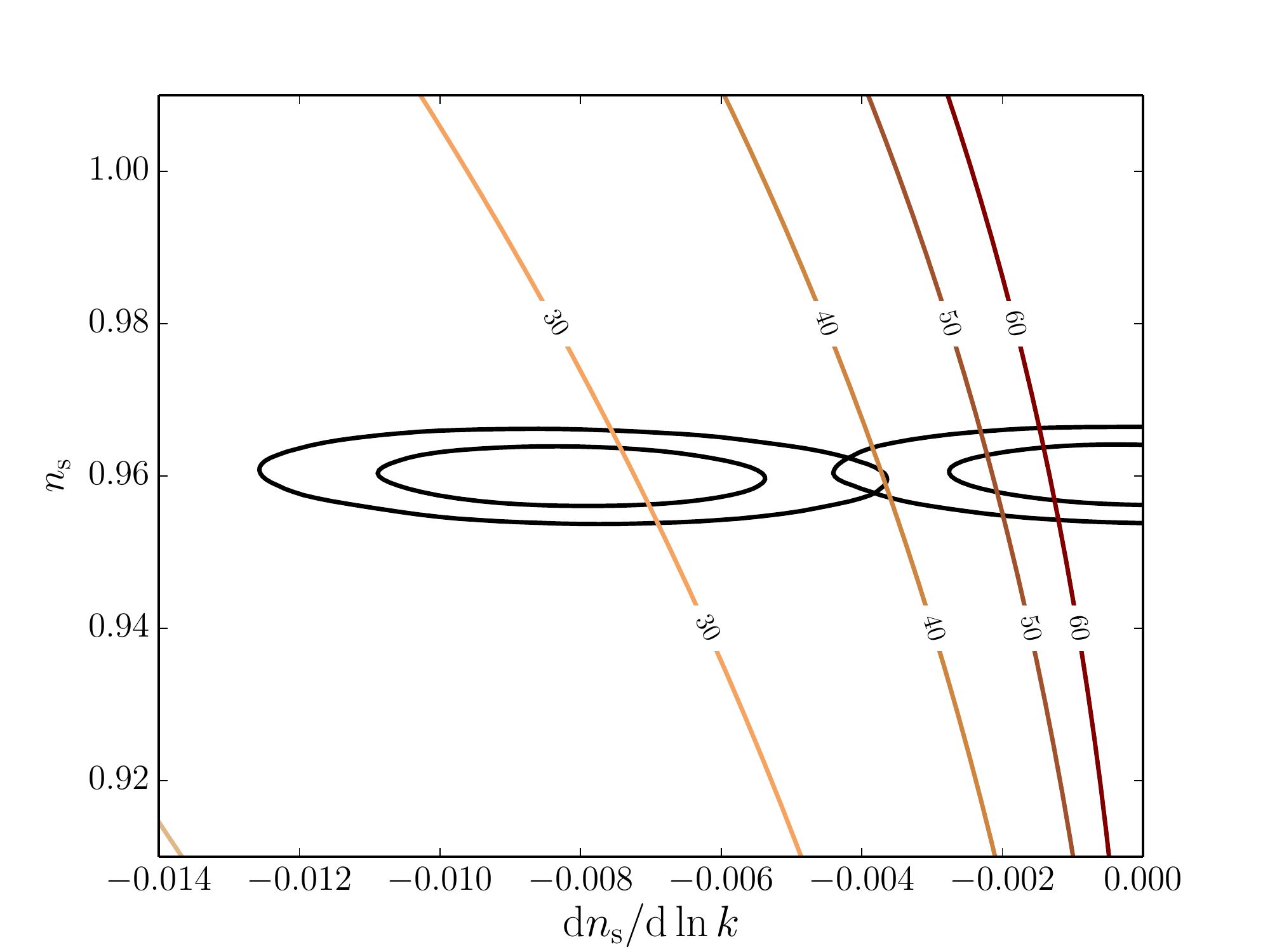}
\caption{Lines of constant $N_e = \{20,30,40,50,60\}$ assuming $r=0$, and the projected 68\%- and 95\%-credible contours derived from the ``csgxcl'' data combination for fiducial models
$(r_{\rm fid}, \mathrm{d}n_\mathrm{s}/\mathrm{d}\ln k_{\rm fid}) =(0,0)$ and $(0,-0.008)$.  \label{fig:slowroll1}}
\end{figure}

We  consider first a fiducial model without tensor perturbations.  Figure~\ref{fig:slowroll1} shows that, given our assumption of $^j\lambda_H = 0$ for $\lambda \geq 3$, a negative running of $\mathcal{O}(10^{-3})$ is required in order to remain in the desired region of $e$-foldings (i.e., $40 \lesssim N_e \lesssim 60$).  However, the projected constraints for a fiducial model with  $\mathrm{d}n_\mathrm{s}/\mathrm{d}\ln k_{\rm fid} = 0$ still comfortably overlap with this region, so even if the true running of the spectral index was zero (and thus our assumption invalid due to inflation lasting for too long), the data would not be able to distinguish the two cases.
If on the other hand the true running should be large and negative, inflation does not continue for long enough to achieve the required number of $e$-foldings.  For $\mathrm{d}n_\mathrm{s}/\mathrm{d}\ln k_{\rm fid} \lesssim -0.008$, the entire 95\%-credible region lies outside the band of sufficient inflation, which means that the data become able to discriminate between models that are consistent with equation~(\ref{eq:H}) and those that are not.\footnote{Note that with current data, the best-fit $\mathrm{d}n_\mathrm{s}/\mathrm{d}\ln k$ is about 50\% larger than this threshold.}  There remains only a small interval, $-0.004 \gtrsim \mathrm{d}n_\mathrm{s}/\mathrm{d}\ln k_{\rm fid} \gtrsim -0.008$, in which the fiducial model is actually inconsistent with the absence of higher-order derivatives due to insufficient inflation, but the data are not sensitive enough to distinguish it from the consistent part of parameter space.

\begin{figure}[t]
\center
\includegraphics[height=.8\textwidth]{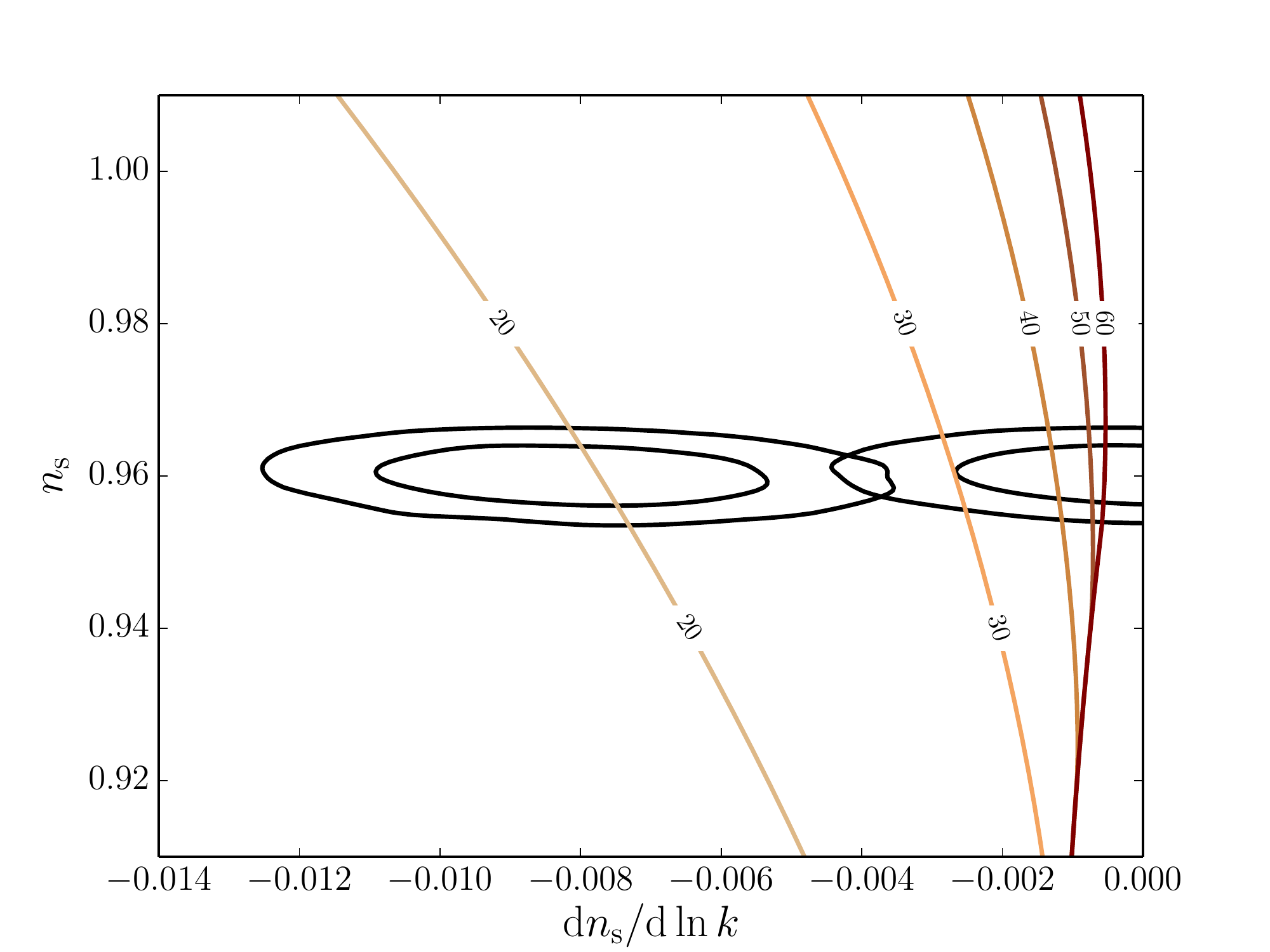}
\caption{Same as figure~\ref{fig:slowroll1}, but for  fiducial models $(r_{\rm fid}, \mathrm{d}n_\mathrm{s}/\mathrm{d}\ln k_{\rm fid}) =(0.1,0)$ and $(0.1,-0.008)$. \label{fig:slowroll2}}
\end{figure}

If the underlying model has $r>0$, the discriminatory power of the data will further improve.  This is demonstrated in figure~\ref{fig:slowroll2} for a tensor-to-scalar ratio of $r=0.1$: the viable region has now migrated closer to zero running and is significantly compressed with respect to the $r=0$ case.  Since $r=0.1$ implies $\epsilon \sim 10^{-2}$, the second order terms in the expression for $\mathrm{d}n_\mathrm{s}/\mathrm{d}\ln k$ in equation~(\ref{eq:phentosloro}) become of the same order as $\xi$;  for small enough $n_\mathrm{s}$ and $\mathrm{d}n_\mathrm{s}/\mathrm{d}\ln k$ close to zero, these terms can even dominate, and force $\xi$ to become negative.  This can cause the inflaton  potential, $V = 3m_\mathrm{Pl}^2/(8\pi) H^2\left(1-3/\epsilon\right)$, to develop a local minimum, thereby trapping the $\phi$-field; 
inflation continues indefinitely under these circumstances, and the lines of constant $N_e$ stack up near the bottom right corner of the figure.   

\section{Conclusions \label{sec:conc}}

In this work we have studied how much the sensitivity to the observational parameters related to inflation can be improved by combining Planck-like CMB measurements with large-scale structure observables from a future large-volume photometric survey such as LSST or Euclid.  Large-scale structure observables constrain length scales significantly
smaller than the reach of CMB measurements, and provide an extended lever arm that is especially useful for measuring the scalar fluctuation spectral index and testing  its possible running.

Using a combination of angular shear power spectrum, angular galaxy power spectrum and cluster mass function, together with Planck-like measurements of the CMB temperature and $E$- and $B$-polarisation, we find that running of the spectral index $\mathrm{d} n_\mathrm{s}/\mathrm{d}\ln k$ can potentially be probed to a precision better than $0.002$ at $1 \sigma$, about a factor of five improvement over current measurements. 
The sensitivity to the spectral index~$n_\mathrm{s}$ likewise sees significant, although less dramatic, improvements: the corresponding $1\sigma$ precision is  $\sim  0.0025$, roughly a factor of three tighter than current constraints.
Not surprisingly, large-scale structure observables offer no direct information on the tensor-to-scalar ratio~$r$.  There is likewise no significant correlation between $r$ and those parameters constrainable by these observables, as indicated by the fact that the sensitivity to~$r$ hardly depends on whether or which large-scale structure data have been included in the analysis.

Importantly, the greatly enhanced constraining power with respect to the running of the spectral index will open a new window into the physics of inflation and test the paradigm of minimal single-field slow-roll inflation.  If $\mathrm{d} n_\mathrm{s}/\mathrm{d}\ln k \sim -0.01$, as hinted at by Planck CMB measurements, then the requirement that inflation lasts long enough will necessitate the presence of non-negligible higher-order derivatives, or multi-field dynamics, which are absent in the simplest inflationary models.

The next generation of large-scale structure surveys will thus truly represent a new milestone in observational cosmology, with the potential to unravel the mysteries of not only the neutrino and dark energy sectors, but also possibly revolutionising our knowledge of the first moments of the universe.

\section*{Acknowledgements}
We acknowledge computing resources obtained via a grant from the Danish e-Infrastructure Cooperation (DeIC).


\bibliographystyle{utcaps}

\bibliography{refs}

\end{document}